\documentclass[structabstract]{aa} 
\usepackage{graphicx}
\usepackage{txfonts}
\begin{document}
   \title{Nainital Microlensing Survey -- detection of short period Cepheids in the disk of M31}


   \author{Y. C. Joshi\inst{1,2}\fnmsep\thanks{yogesh@aries.res.in},
D. Narasimha\inst{3},
A. K. Pandey\inst{1},
R. Sagar\inst{1}
          }
   \institute{
Aryabhatta Research Institute of Observational Sciences (ARIES), Manora peak, Nainital, India	      
\and
Astrophysics Research Centre, School of Mathematics \&\ Physics, Queen's University, Belfast, BT7 1NN, UK
\and
Tata Institute of Fundamental Research, Homi Bhabha Road, Mumbai, India
}
   \date{Received ; accepted }

 
  
\abstract
{Cepheids are the primary distance indicators for the external galaxies and
discovery of large number of Cepheid variables in far-off galaxies offers a unique
opportunity to determine the accurate distance of the host galaxy through their
period-luminosity relation.} 
{The main purpose of this study is to identify short-period and relatively faint Cepheids
in the crowded field of M31 disk which was observed as part of the Nainital
Microlensing Survey.}
{The Cousins $R$ and $I$ band photometric observations were carried out in the direction
of M31 with an aim to detect microlensing events. The data was obtained with a 1-m telescope
on more than 150 nights over the period between November 1998 to January 2002. The data was
analysed using the pixel technique and the mean magnitudes of the Cepheids were determined
by correlating their pixel fluxes with the corresponding PSF-fitted photometric magnitudes.}
{In the present study we report identification of short-period
Cepheid variables in the M31 disk. We present a catalogue of 39 short-period ($P < 15$ days)
Cepheids in a $\sim13'\times 13'$ region of the M31 disk and give positions and pulsation
periods along with their $R$ and $I$ bands photometric magnitudes wherever possible. Most
of the Cepheids are found with $R$ (mean) $\sim$ 20 -- 21 mag and the dense phase coverage
of our observations enabled us to identify Cepheids with periods as short as 3.4 days.
The period distribution of these Cepheids peaks at $logP \sim 0.9$ and 1.1 days.}
{We demonstrate that using pixel method, faint and short-period Cepheids in M31 can be
detected even with small-size telescopes and moderate observing conditions.}

\authorrunning{Y. C. Joshi et al.}

\titlerunning{Detection of short period Cepheids in the disk of M31}

\maketitle
\section{Introduction}
In the past decade, the Andromeda galaxy (M31) has been a target of search
for gravitational microlensing events by several wide-field surveys e.g. AGAPE
(Ansari et al. 1997, 1999), Columbia-VATT (Crotts \& Tomaney 1996), POINT-AGAPE
(Auri\'{e}re et al. 2001, Paulin-Henriksson et al. 2003), WeCAPP (Riffeser et al. 2001),
MEGA (de Jong et al. 2004) and Angstrom (Kerins et al. 2006). To discriminate
microlensing events from  known types of variable stars, these surveys need
continuous observations for a long time span with good temporal sampling though
for a short period of time each night. Such observations are therefore perfectly
suited to the detection of  variable stars (e.g. Cepheids, Miras) and optical transient
events. Several groups dedicated to search for microlensing events in the direction
of M31 have already uncovered a large number of variable stars, most of which are
previously unidentified (Joshi et al. 2003, Ansari et al. 2004, An et al. 2004,
Fliri et al. 2006) as well as nova outbursts (Joshi et al. 2004, Darnley et al. 2004)
as a major by-product.These variable stars are of cosmological interest,
particularly Cepheids which are massive ($M \sim 3-20 M_\odot$) pulsating stars
placed in the instability
strip of the Hertzsprung-Russell diagram. They can be identified by their characteristic
`saw-tooth' shaped light curves and large intrinsic brightness. The correlation of the
period of pulsation with their intrinsic magnitudes makes them useful to measure distances
to galaxies in the Local Group and nearby clusters of galaxies.
In recent times, a substantial work has been done on the determination of the distance of
M31 using population I Cepheids through their period-luminosity relation, with a broad
range of distances (730 to 790 kpc) reported by different groups (e.g. Freedman et al. 2001,
Joshi et al. 2003, Vilardell et al. 2007).

Starting in November 1998, we undertook a long-term project, the `Nainital Microlensing
Survey' to search for microlensing events towards M31 using a 1-m telescope in Nainital,
India. The survey has good temporal coverage during September/October to January for four
consecutive observing seasons of M31 and offers an excellent opportunity to search for
variable stars and optical transients in the disk of M31. In our earlier survey papers,
we reported variable stars including long-period Cepheids and irregular variables
(Joshi et al. 2003, hereafter referred as JOS03), classical novae (Joshi et al. 2004) and
a microlensing candidate event (Joshi et al. 2005). In this paper we present a catalog of
short-period Cepheids ($P < 15$ days) detected in the survey. While a detailed
description of our observations and reduction can be found in JOS03, a brief overview is
given in Sects. 2. In Sect. 3 we describe the pixel analysis procedure used to identify
Cepheids in our data. The detection  procedure of Cepheids and their catalogue are given
in Sect. 4 and 5 respectively.  Our results are discussed in Sect. 6.

\section{Observations and data reduction}
Cousins $R$ and $I$ band photometric observations of the target field centered at
$\alpha _{2000}$ = $0^{h} 43^{m} 38^{s}$; $\delta_{2000}$ = $+41^{\circ}09^{\prime}.1$,
were obtained with a 1-m Sampurnanand Telescope at the Manora Peak, Nainital, India.
The total integration  time during our survey ranges from $\sim$ 30 minutes to 2
hours each night and a median seeing during the observations was $\sim$ 2.2 arcsec.
Due to time constraints, it was not possible to observe the target field in both the
filters each night so we put an observing priority on the $R$ band. Images observed in
poorer seeing than 3.5 arcsec were removed from our analysis to avoid blending problems
since a large number of stars are present in the target field. In the 4 years long observing
run,  we were finally left with a total of 133 nights data in $R$ band and 115 nights
data in $I$ band with a total time span of $\sim$ 1200 days. In addition, we also observed
the Landolt's standard field SA98 (Landolt 1992) on the photometric night of 25/26 October
2000 in order to derive the transformations equations to standard magnitudes. A log of
observations in the electronic form is given in JOS03.

The basic steps of image processing which include bias subtraction,
flat fielding, masking of bad pixels and cosmic ray removal were performed using
IRAF\footnote{Image Reduction and Analysis Facility (IRAF) is distributed by the National
Optical Astronomy Observatories, which are operated by the Association of Universities for
Research in Astronomy, Inc., under cooperative agreement with the National Science Foundation.}.
In order to improve the signal-to-noise ratio, all images in a particular passband were
combined on a nightly basis resulting in a single image per filter per night.

Stellar photometry of all the images in both the filters were carried out for about 4400
resolved stars using DAOPHOT photometry (Stetson 1987). PSF was obtained for each frames
using 25-30 relatively bright uncontaminated stars. The DAOPHOT/ALLSTAR routine was used to
calculate the instrumental magnitude of these stars. The absolute calibration has been done
using Landolt's (1992) standard field SA98.
The typical photometric error was estimated to be about 0.04 mag for stars at $R$ = 20 mag,
increasing to 0.20 mag at $R$ = 22 mag.

\section{Image analysis using the pixel method}
Since our target field of M31 is composed of largely faint stars, neither all the stars
are well resolved nor each variable star is sufficiently bright at minimum brightness to
obtain reliable DAOPHOT photometry.
The  incompleteness in our data set begins at $R \sim$20 and $I \sim$19.5 - this is precisely
the brightness that we would expect of the lower luminosity short-period Cepheid variables.
Therefore, in the present study, we used the pixel method to analyse our data
which relies on the monitoring of pixel light curves and their shape analysis. This method was
originally proposed by Baillon et al. (1993) and implemented by Ansari et al. (1997) and
others. In the pixel method if a star of flux $F_{star}$ is increased, either due to intrinsic
variability or gravitationally lensed, then by subtracting the original flux from the amplified
flux of the star, one gets an increase in flux equal to $(A-1)F_{star}$ above the photon noise
where $A$ is the flux amplification of the star. Thus by following $\Delta F$ with time, we in
fact monitor the variation in the flux of the target star.

Our implementation of pixel method is described in detail in  Joshi et al. (2005) where
we report detection of the first microlensing candidate in our survey. In brief,
to implement the pixel method in our data, we first chose a reference frame taken in good
photometric conditions with low sky background and relatively good seeing ($\sim 1^{''}.5$).
Images were normalized with respect to the reference frame in the following three steps.
\begin{enumerate}
\item We geometrically aligned all the images with better than $\pm$0.05 arcsec accuracy through
rotation and shifting with respect to the reference frame.
\item The photometric conditions were different on different nights during our observing
runs which we corrected by normalizing all the images with respect to the median background
of the reference frame.
\item To further reduce the fluctuations due to the seeing problem, we constructed  a
{\it superpixel} of $7\times7$ pixels ($\sim 2.5\times2.5$ arcsec$^2$) of which combined
flux is given by
\begin{equation}
\phi_{superpixel}(i,j) = \sum_{k=i-3}^{i+3} \sum_{l=i-3}^{i+3} \phi_{pixel}(k,l)
\end{equation}
where $\phi_{pixel}(i,j)$ is the pixel flux at any pixel coordinate $(i,j)$.
\end{enumerate}
In our subsequent discussion, we use the term {\it pixel} for convenience when referring to
the superpixel. After corrections,
the photon counts in any pixel are expected to only exhibit a flux
variation ($\Delta F$) above the background level if any star or stars falling over the pixel
show intrinsic brightness variations. It is worth mentioning here that to detect any variation
in the flux, this change must be significantly above the background level. In our analysis,
we have considered flux variation as significant if at least 3 consecutive points were above
$3 \sigma$ level in each observing season.
It is important to remark that the pixel method is more sensitive to the detection of faint
but large amplitude than to bright but low amplitude variables (An et al. 2004).

\section{Identification of Cepheid variables}
While a substantial number of M31 Cepheids are reported in the
long-period range (P $\sim$ 7-60 days), short-period Cepheids are not well reported as they
are relatively faint and have variations of smaller amplitude.
In the present study, we use our data to search for short-period Cepheids ($P<15$ days). To
identify these variables, we first masked all the bright stars ($R <$ 19.5 mag ) in a 10 pixel
radius ($\sim$ 2$\times$FWHM) in all the frames. The remaining pixels are searched for
variability in their light curves and we identified few thousands pixels in our target field
which were further analysed for their periodic variations.

\subsection{Period determination}
Following JOS03, we used a modified version of the Press \& Rybici (1989) FORTRAN program
based on the method of Horne \& Baliunas (1986) to determine the period of the variable stars.
This method uses a series of sinusoidal signals to best match the time series, and hence find the
period. The time series is convolved with sinusoidal curves until the peak of the convolution
is found. On average, we have an $R$ band image every 3 days (with wide gaps between observing
seasons), thus we started searching for variations with a minimum period of 3 days and with an
increment of 0.01 days. As this data has already been used to find longer period Cepheid variables
using DAOPHOT profile fitting techniques (see JOS03), we searched for periods up to 15 days.

\subsection{Selection criteria to identify Cepheids}
A systematic search for  variable stars in the data was performed by determining the shape of the
periodic variations in the selected pixels. The $R$ band images with their improved temporal coverage
and photometric accuracy, were used  to characterize the period variations in terms of mean brightness,
period and amplitude of the Cepheids (the $I$ band data was not used at this
stage).  Measurements
were flagged bad when the error in the pixel flux was more than 400 ADU  and discarded from further
analysis. We initially only used those pixels which follow following criteria:
\begin{enumerate} 
\item After the bad pixels rejection, individual pixel flux measurements are available
for at least 50\% of the total nights.
\item The pixel flux shows a periodic variation with a period less than 15 days 
\item The amplitude of any periodic flux variation is greater than 200 ADU in the $R$ band data.
\end{enumerate}
We thus shortlisted 1177 pixels which passed these criteria.
Using the period determined in the previous section we derived the phase for each
observation.
Since there are large errors in the individual pixel fluxes, we binned the data
in 20 bins of width 0.05 in phase. For those bins having measurements, the weighted
mean pixel flux and error were determined. The binned light curves have a smaller
scatter, allowing a better visual identification of Cepheid-like variability. A
total of 39 Cepheids with periods ranging from $\sim$ 3.5 to 15 days were identified.
The $I$ band data was then binned discarding those pixels having flux errors
larger than 600 ADU. We phased the $I$ band data using the same period as estimated
through $R$ band and found the periodic variations in all the Cepheids but NMS-M31V05.
However, we still consider NMS-M31V05 as a Cepheid variable as shape of its $R$ band
phase light curve looks convincing.
 
\begin{figure*}
\centering
\includegraphics[width=18cm,height=20cm]{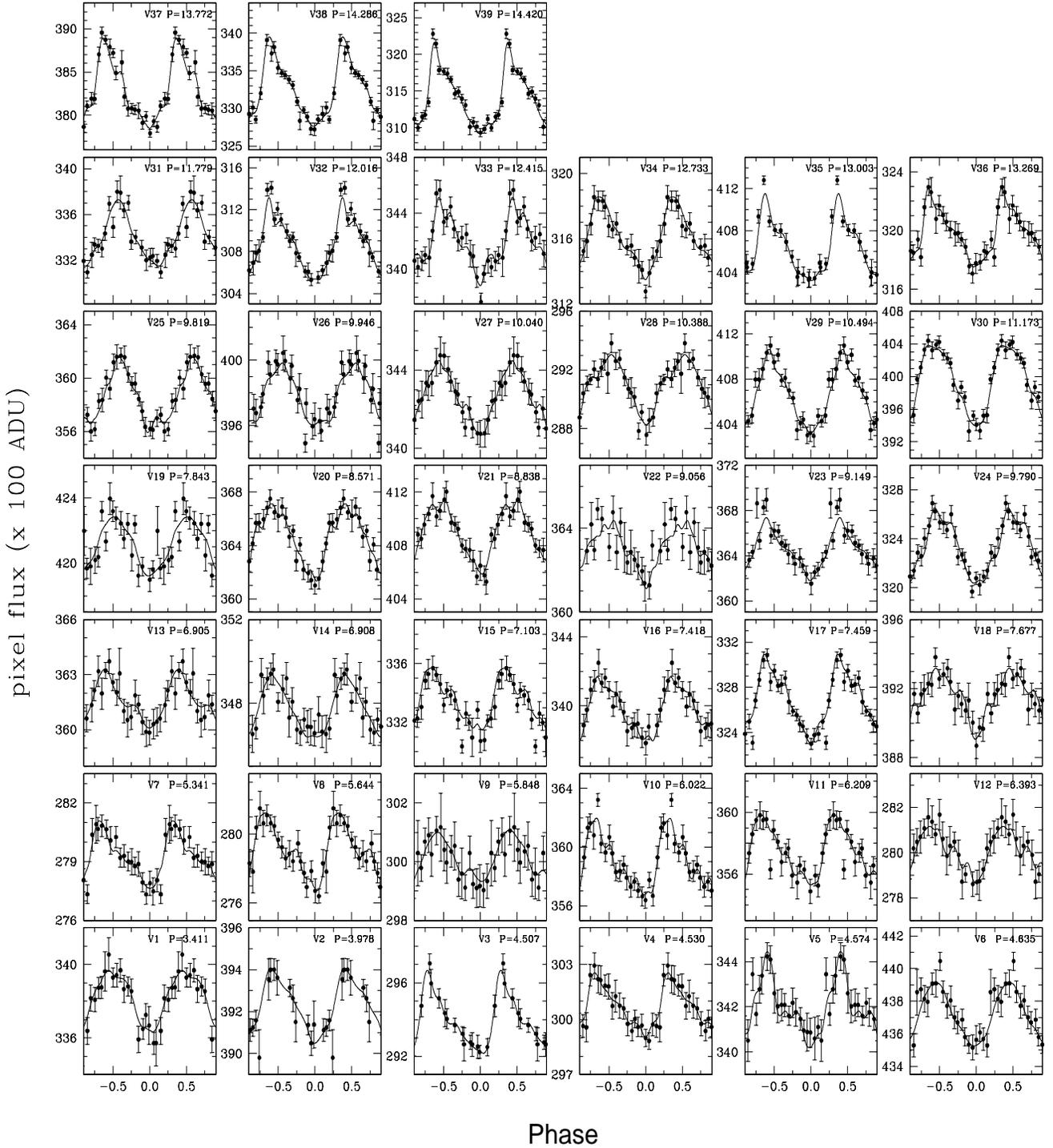}
\caption{$R$ band phase light curves of the 39 Cepheids. Periods of the Cepheids are
given at the top of each light curve. Phase is plotted twice and in such a way
that the minimum flux falls near to zero phase.
We use GNUPLOT {\it acsplines} routine to interpolate the light curves which
approximates the data with a `natural smoothing spline'. We have not used statistical
errors for the weighting and instead a constant value was used as smoothing weights.}
\label{cephR}
\end{figure*}
\begin{figure*}
\centering
\includegraphics[width=18cm,height=20cm]{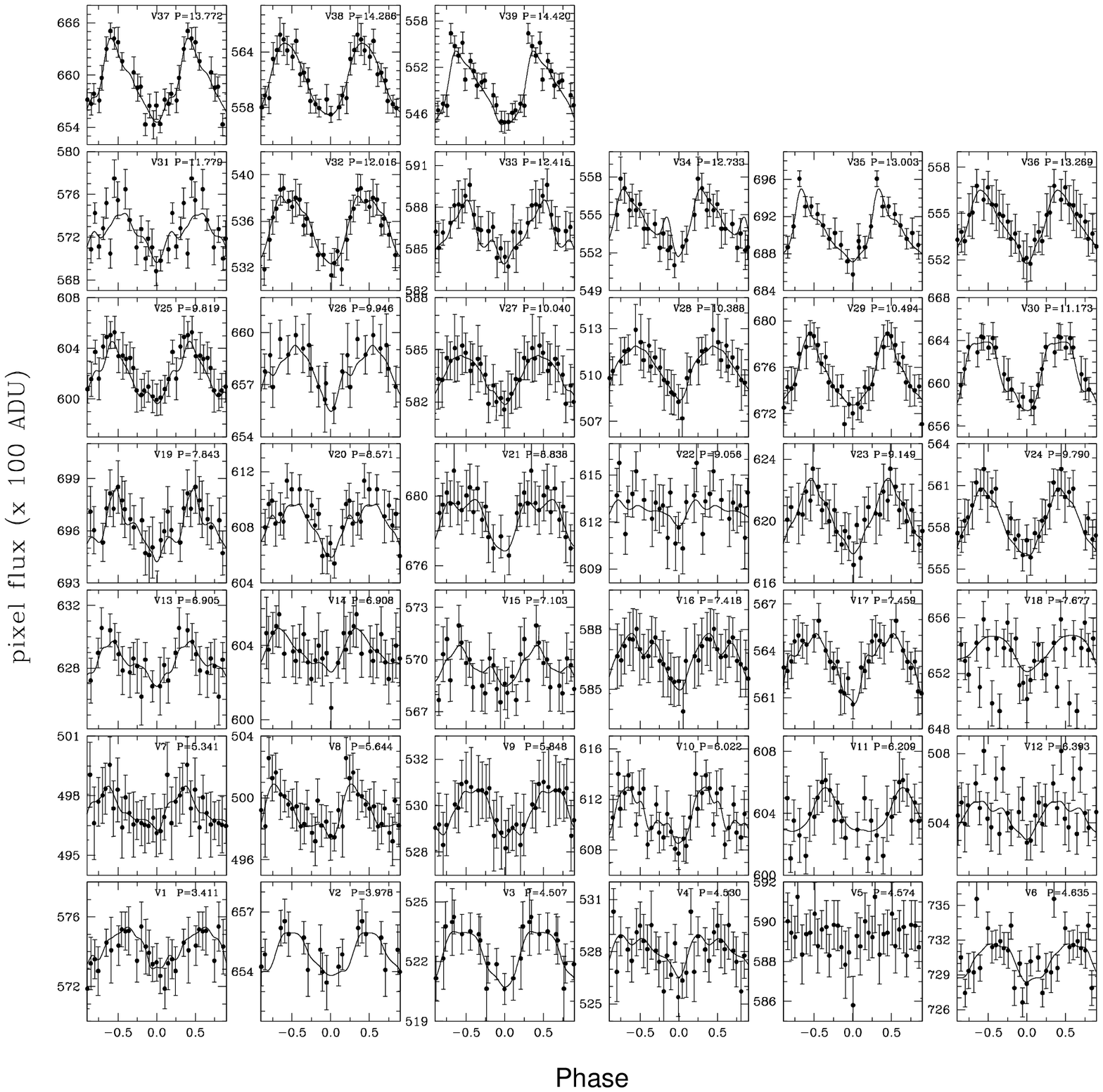}
\caption{Same as in Fig. 2 but for the $I$ band.}
\label{cephI}
\end{figure*}

In Fig.~\ref{cephR} and Fig.~\ref{cephI}, we show the light curves of 39 detected
Cepheids in $R$ and $I$ bands respectively in order of increasing period. The period
of each Cepheid is given at the top of its pixel light curve.
Given the lesser observations in $I$ band, the $I$ band light curves have larger
uncertainties than their $R$ band counterparts, particularly in the extremely short-period
regime which contains faint stars and exhibit low-amplitude variability. 

\subsection{Determination of mean magnitude}
To determine the mean magnitude of Cepheids, we first calculate phase-weighted mean flux
as
\begin{equation} 
F_{mean} = 0.5\mathrm{\sum_{i=1} ^{n}}(\phi_{i+1}-\phi_{i-1})F_{pixel}
\end{equation}
where n is the total number of observations, $\phi_{i}$ is the phase of
$i^{th}$ observation in order of increasing phase after folding the period.
The equation requires non-existent entities $\phi_{0}$ and $\phi_{n+1}$
which is set identical to $\phi_{n}$ and $\phi_{1}$ respectively. 

Unlike in other pixel surveys (e.g. An et al. 2004, Fliri et al. 2006) where pseudo-magnitudes
of the variable stars were determined from their flux variations, we calculated the absolute
magnitude of the Cepheids using their photometric magnitudes if available through
DAOPHOT photometry.  To do this, whenever we identified a Cepheid-like light curve in the
pixel method, we forcefully run IRAF DAOPHOT `FIND' routine around that pixel position in all
the frames of both the filters. We did not consider a star identified if it lies more than
3 pixels ($\sim$ 1 arcsec) different from the
given pixel coordinates as it could  be a different star or may have been the result of a 
blending problem. We did not find all the stars in all the images because most of
these Cepheids were faint enough to be close to our detection limit and partly due to
observations being held in different sky conditions over the 4 years period. In the next
step, we carried out PSF photometry around these stars to estimate their precise photometric
magnitude and did not use  any selection criteria to accept the magnitude unlike in JOS03.
We found more than 30 photometric measurements for most of the Cepheids but could not find
reliable photometric magnitudes for 6 Cepheids in $R$ band and 15 in $I$ band.

\begin{figure}
\centering
\includegraphics[width=9.0cm,height=7.4cm]{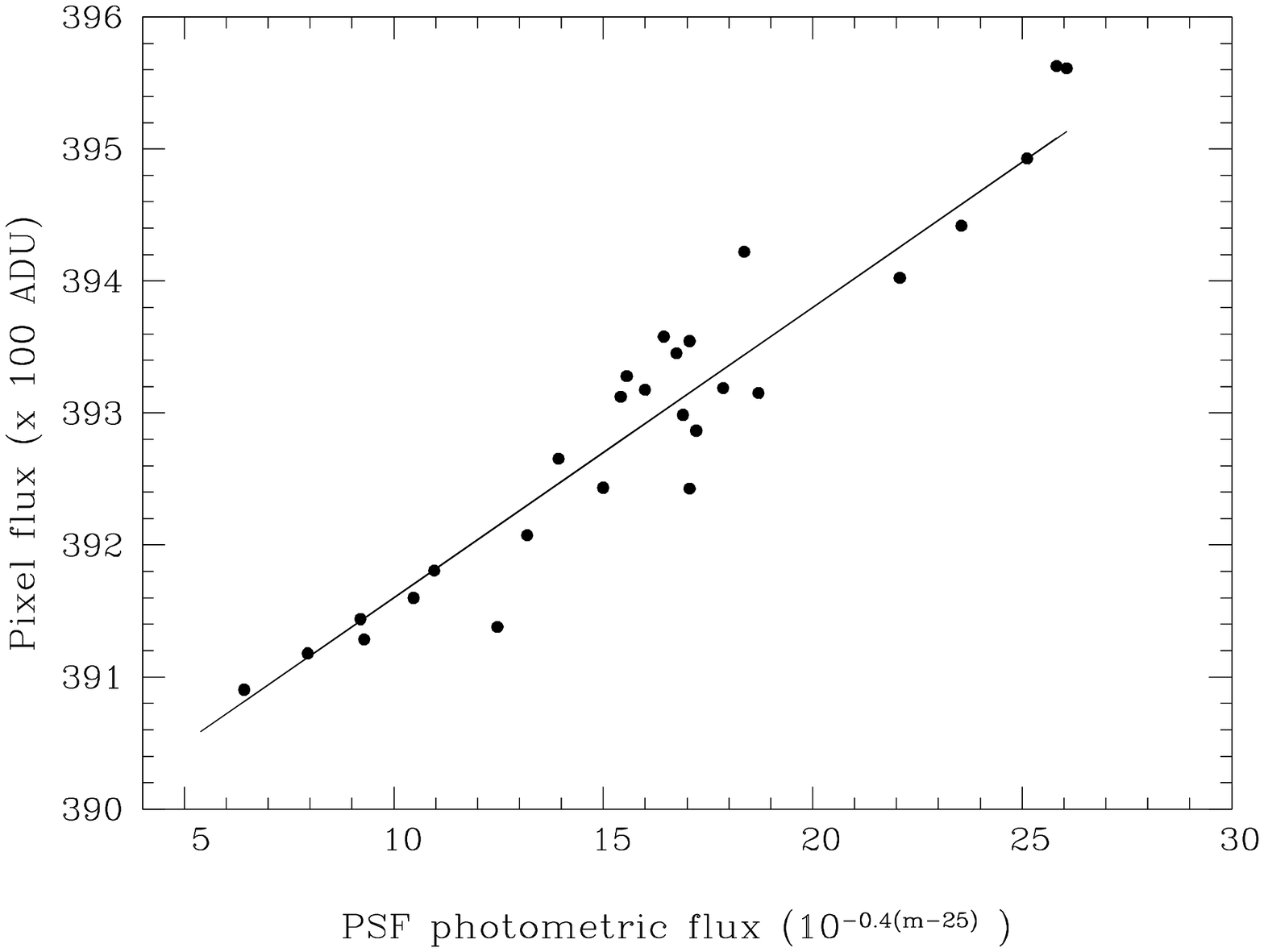}
\vspace{-2.3cm}
\caption{A correlation between photometric fluxes determined from PSF-fitted photometric
magnitudes and corresponding pixel fluxes of a Cepheid NMS-M31V2 identified in our survey.}
\label{ceph_comp}
\end{figure}

For any Cepheid, we correlated photometric magnitudes with their corresponding pixel
fluxes to determine the background flux, which varied with spatial
positions due to the large background gradient in the M31 disk. We note here that
all the images used to determine the pixel fluxes are background corrected and, therefore,
contains the same background level at the position of the Cepheids. We derived a linear
relation between the PSF-fitted photometric magnitudes converted into photometric fluxes and
that of the pixel fluxes. We kept the slope fixed for all the Cepheids and neglected the colour
term in the transformation as the error in our $I$ band photometry was expected to be larger
than that from the colour term itself. In correlating the two fluxes, we derived the background
flux only when the two data were correlated to $>$80\%. For example, Fig.~\ref{ceph_comp} shows
the correlation between PSF-fitted photometric fluxes and pixel fluxes for a Cepheid NMS-M31V2.
This star was identified in only 41 nights on the DAOPHOT photometric identification as star was
fainter than 21 mag in $R$ band, even at its maximum brightness. We used 28 nights after three
iterations of one-sigma clipping to determine the background flux at the position of this Cepheid.
In a similar way, we determined the background flux for each Cepheid at their pixel positions.
Using the fixed slope and background flux for each Cepheid, we converted the phase weighted mean
pixel flux into the mean magnitude for all 33 Cepheids in $R$ and 24 Cepheids in the $I$ band.
The standard deviations in our magnitudes could be as much as 0.20 mag in $R$ and 0.30 mag
in the $I$ band. The main source of error in the magnitudes is dominated by the transformation
from pixel flux to photometric magnitude due to the lack of precise photometry towards the fainter
end. Note that most of these Cepheids are observed close to the detection limit of our telescope,
some of them may not even be detected in their minimum brightness phase.

\subsection{Astrometry}
Astrometry is performed on one of the best image obtained on a photometric night
with relatively good seeing ($\sim 1^{''}.7$) and lower sky background. To convert the pixel
coordinates $(X,Y)$ into celestial coordinates ($\alpha, \delta$), reference positions
of 324 bright stars from the USNO catalogue (Monet et al. 2003) were used to find
linear astrometric parameters. The pixel positions of the detected Cepheids were then
converted to the J2000 celestial coordinates in the equatorial system using the IRAF
tasks of {\it ccmap} and {\it cctran}. The coordinates matches within
$\sim$ 1.0 arcsec with those given in the Magnier catalogue (Magnier et al. 1992)
and should be considered to be the typical inaccuracy in our astrometry.

\begin{table*}
\centering
\caption{A list of 39 short-period Cepheids identified in the present study with their
characteristic parameters. A cross-reference ID and period of the common Cepheids found
in the POINT-AGAPE catalogue (An et al 2004, suffixed as PA04) are given in columns
8 and 9 respectively. Common Cepheids identified by DIRECT (Kaluzny et al. 1999), Joshi et
al. (2003) and WeCAPP (Fliri et al. 2006) are given in the last column with prefixes D, J
and W respectively. Their corresponding periods are also mentioned in the brackets.}
\begin{tabular} {cccccccccl}
\hline
Cepheid  & RA (J2000) & DEC (J2000) & P    &$\overline{R}$&$\overline{I}$& $A_R$ & PA04 & P & Other Identification   \\
(NMS-)   & hh:mm:ss   & dd:mm:ss    &(Days)&   (mag)      & (mag)        & (mag) &  ID  & (Days) &  \\
\hline
M31V1  & 00:43:48.20 & 41:12:55.9 &  3.411$\pm$0.001 &   -     &   -   &  -    &       &       &    \\
M31V2  & 00:43:11.14 & 41:10:30.8 &  3.978$\pm$0.001 & 22.29   &   -   & 0.26  &       &       &    \\
M31V3  & 00:44:02.59 & 41:11:37.5 &  4.507$\pm$0.001 & 21.08   &   -   & 0.57  & 69301 & 4.508 &     \\
M31V4  & 00:43:20.94 & 41:04:07.7 &  4.530$\pm$0.001 & 20.84   &   -   & 0.18  &       &       &    \\
M31V5  & 00:43:30.24 & 41:10:35.6 &  4.574$\pm$0.001 & 21.52   &   -   & 0.31  & 72533 & 4.634 &       \\
M31V6  & 00:43:28.65 & 41:14:53.2 &  4.635$\pm$0.001 & 20.92   &   -   & 0.22  & 72356 & 4.581 &    \\
M31V7  & 00:43:54.95 & 41:08:12.3 &  5.341$\pm$0.002 & 21.26   & 20.83 & 0.27  & 69645 & 5.346 &    \\
M31V8  & 00:43:59.13 & 41:08:07.8 &  5.644$\pm$0.002 & 20.78   & 20.06 & 0.22  & 69680 & 5.636 &    \\
M31V9  & 00:43:19.72 & 41:05:32.4 &  5.848$\pm$0.004 &   -     &   -   &  -    &       &       &    \\
M31V10 & 00:43:20.88 & 41:10:24.5 &  6.022$\pm$0.002 & 21.18   & 20.60 & 0.37  & 74876 & 6.026 & W2583 (6.021, 6.021)	\\
M31V11 & 00:43:21.70 & 41:08:19.4 &  6.209$\pm$0.003 &   -     &   -   &  -    & 75216 & 6.209 &    \\
M31V12 & 00:43:32.84 & 41:04:53.5 &  6.393$\pm$0.004 &   -     &   -   &  -    & 73716 & 6.397 &    \\
M31V13 & 00:43:14.03 & 41:09:24.9 &  6.905$\pm$0.004 & 20.67   & 19.70 & 0.13  &       &       & W1314 (6.909, 6.906)	\\
M31V14 & 00:43:21.06 & 41:08:14.1 &  6.908$\pm$0.004 & 20.59   &   -   & 0.10  &       &       & W98 (6.908, 6.899)    \\
M31V15 & 00:43:45.28 & 41:12:21.3 &  7.103$\pm$0.003 & 20.74   & 20.11 & 0.21  &       &       &    \\
M31V16 & 00:43:21.57 & 41:08:02.4 &  7.418$\pm$0.004 & 20.74   &   -   & 0.18  & 75465 & 7.413 &    \\
M31V17 & 00:43:43.71 & 41:11:48.9 &  7.459$\pm$0.002 & 20.55   &   -   & 0.28  & 71096 & 7.464 & D883 (7.459), J01(7.459) \\
M31V18 & 00:43:35.44 & 41:15:05.2 &  7.677$\pm$0.004 & 20.41   & 20.34 & 0.13  &       &       &    \\
M31V19 & 00:43:17.49 & 41:12:11.3 &  7.843$\pm$0.004 & 20.60   & 19.89 & 0.15  & 74607 & 7.852 & W5037 (7.842, 7.849)	\\
M31V20 & 00:43:23.23 & 41:10:25.3 &  8.571$\pm$0.004 & 20.16   & 19.67 & 0.17  & 74753 & 8.551 & W2562 (8.567, 8.572), J02(8.566)\\
M31V21 & 00:43:28.04 & 41:13:55.4 &  8.838$\pm$0.003 & 20.71   & 20.46 & 0.27  & 72015 & 8.831 &  J03(8.836)  \\
M31V22 & 00:43:45.37 & 41:15:09.7 &  9.056$\pm$0.012 &   -     &   -   &  -    &       &       &    \\
M31V23 & 00:43:44.82 & 41:15:01.0 &  9.149$\pm$0.009 & 20.37   & 19.65 & 0.20  &       &       & D1219 (9.173), J04(9.160)     \\
M31V24 & 00:43:53.27 & 41:12:46.1 &  9.790$\pm$0.004 & 20.58   & 20.10 & 0.25  & 70319 & 9.772 & D2879 (9.790), J05(9.790)     \\
M31V25 & 00:43:33.65 & 41:11:52.9 &  9.819$\pm$0.005 & 20.76   &   -   & 0.26  & 72649 & 9.550 &    \\
M31V26 & 00:43:38.79 & 41:15:53.8 &  9.946$\pm$0.009 & 20.14   & 19.83 & 0.11  &       &       &    \\
M31V27 & 00:43:40.67 & 41:12:44.8 & 10.040$\pm$0.004 & 20.93   & 20.64 & 0.23  & 69993 & 10.023 &    \\
M31V28 & 00:43:30.49 & 41:03:36.4 & 10.388$\pm$0.006 & 20.53   & 20.27 & 0.20  & 87421 & 10.375 & J06(10.383)	\\
M31V29 & 00:43:29.67 & 41:14:12.0 & 10.494$\pm$0.004 & 20.71   & 20.35 & 0.34  & 72289 & 10.495 & J07(10.500)  \\
M31V30 & 00:43:00.01 & 41:08:33.3 & 11.173$\pm$0.006 & 20.04   & 20.25 & 0.27  &       &	& W490 (11.168, 11.172), J08(11.19)  \\
M31V31 & 00:43:51.25 & 41:14:24.0 & 11.779$\pm$0.013 & 20.95   & 20.26 & 0.34  &       &	&    \\
M31V32 & 00:43:21.70 & 41:05:02.4 & 12.016$\pm$0.009 & 20.49   & 20.06 & 0.29  & 75721 & 12.050 &    \\
M31V33 & 00:43:28.98 & 41:10:12.8 & 12.415$\pm$0.008 &   -     &   -   &  -    & 72505 & 12.417 &    \\
M31V34 & 00:43:50.98 & 41:12:56.9 & 12.733$\pm$0.005 & 21.81   & 20.85 & 0.74  & 70598 & 12.706 &    \\
M31V35 & 00:43:35.09 & 41:15:32.0 & 13.003$\pm$0.007 & 20.74   & 19.56 & 0.36  &       &	&    \\
M31V36 & 00:43:46.67 & 41:11:30.0 & 13.272$\pm$0.007 & 20.60   & 20.14 & 0.21  & 71271 & 13.274 &    \\
M31V37 & 00:43:26.25 & 41:12:01.6 & 13.770$\pm$0.006 & 20.26   & 19.68 & 0.31  & 72459 & 13.772 & J09(13.773)	 \\
M31V38 & 00:43:47.94 & 41:10:02.6 & 14.286$\pm$0.006 & 19.53   & 19.09 & 0.16  & 71168 & 14.256 &    \\
M31V39 & 00:43:42.97 & 41:10:17.6 & 14.420$\pm$0.005 & 20.86   & 19.97 & 0.56  & 70712 & 14.454 & J10(14.420)  \\
\hline										      
\end{tabular}									    
\end{table*}									    

\section{The Catalogue}
Using the pixel method as a detection technique for the faint Cepheids, we have
significantly increased the number of Cepheids in our target field of M31.
A list of 39 short-period Cepheids identified in our survey is given in Table 1. The Cepheids are
sorted in order of increasing period. The table contains the object identifier,
right ascension (RA), declination (Dec), period ($P$), $R$ and $I$ band phase weighted mean
magnitudes ($\overline{R}$, $\overline{I}$), and amplitude
of the pulsation in the $R$ band ($A_R$). The objects are assigned names in  the format
NMS-M31Vn where n is the Cepheid sequence number and acronym NMS is used for Nainital
Microlensing Survey. Celestial coordinates are given for J2000. Whenever any Cepheid is reported
as a variable star in the POINT-AGAPE survey catalogue (An et al. 2004), we give their
identification number and period in the columns 8 and 9. If any other references
are found corresponding to the Cepheid identified in our study (see the discussion in Sect. 6.1),
we also give those identifications and periods in the last column of the table.

In the present study, we detected all the 10 Cepheids in 7-15 days period range which were reported
in JOS03. However, we identified 15 new Cepheids in the same period range. This was possible
mainly due to our approach where we search for pixel variability around each pixel instead of
identifying stars, and subsequently looking for variability around them after filtering
through various selection cuts. On some brighter Cepheids (e.g. M31V38), the smaller
amplitude could be due to their binary nature or may be significantly blended by nearby
bright stars that is unresolved in our observations.

The PSF FWHM of the images used in our analysis varies from 1.5 -- 3.5 arcsec, which is
equivalent to about 5 to 12 pc at the distance of 780 kpc. This indicates that despite
large intrinsic brightness of the Cepheids, these stars are very much likely to be
affected by the flux contribution of hundreds of other neighbouring stars in our target
field (i.e., blended), which can significantly increase their observed magnitude and
decrease the amplitude.
Using High-resolution HST images, Mochejska et al. (2000) concluded that the average
flux contribution from the bright companions that are not resolved on the ground-based
images is about 19\% of the flux of the Cepheid in $V$ band.  
Macri et al. (2006) pointed out that fainter and low amplitude Cepheids like those detected in
our pixel survey are more affected by this blending problem. In general, different estimates
put a blending of 0.1 to 0.3 mag in $B$ and $V$ bands in M31 Cepheids (Mochejska et al.
2000, Kiss \& Bedding 2005, Vilardell et al. 2007 and references therein). In the present case
it could be even larger due to choice of our filters ($R$ and $I$ bands) since large
numbers of red stars are present in M31. We therefore caution readers that the magnitudes
and amplitude for the Cepheids given in Table 1 should not be considered definitive
and much more precise photometry is needed to accurately determine these values.

One major problem of our analysis is the uncertainty in the determination of the colour
$(R-I)$, which is, unfortunately, mainly caused by the poor light-curves and smaller phase
sampling in the $I$ band, as well as the blending problem. Therefore, we are not in a position
to discuss the colour-magnitude diagram of these Cepheids.

\section{Discussion}
\subsection{Comparison with other catalogues}
The list of Cepheids reported here is compared with those surveys having identified
variable stars within our relatively small field of view. A comparison with the $V$-band
photometric observations of Magnier et al . (1997)
and $BVI$ photometric observations carried out by the DIRECT project (Kaluzny et al. 1999)
has already been done in JOS03. No additional short-period Cepheids in common have been
found, since all of then were already previously identified. 

The WeCAPP microlensing survey recently reported a catalogue of
23781 variable stars including 33 population I Cepheids and 93 population II Cepheids
(Fliri et al. 2006). Though WeCAPP survey mostly observed towards the bulge of M31,
only a small portion of our target field was common with them in which they found
11 Population I Cepheids. However, we identified only 6 of them in our survey as rest
of them were too faint to be detected in our photometry.
For each Cepheid, WeCAPP survey has given two
different periods in $R$ and $I$ bands and we found our periods agree to within 0.001 day with
one of their two periods given for any Cepheid (see, column 10 of Table 1).

The most complete list of variable stars in M31 to date is given by the POINT-AGAPE microlensing
survey which has produced an exhaustive list of 35414 variable stars in M31 (An et al. 2004).
It is worth mentioning here that POINT-AGAPE survey has not characterized their variable
stars as Cepheids and just listed them as the variables in their catalogue.
Two of their fields (7 and 8) fall in our target field. We found 25 Cepheids in the present
study which were also listed in their catalogue. The celestial coordinate of some of the
common variables between two surveys are separated by as much as 4 arcsec but found with
almost the same period. This could represent the typical combined astrometric inaccuracy between
these two catalogues. Furthermore, some of their variable stars were found less than 5 arcsec
away from each other. For example, one of our Cepheid NMS-M31V8 matches with two of their variable
stars with ID number 69654 and 69680 which lie at a separation of just 2 arcesc but identified
with vastly different period of 171.791 and 5.636 days respectively in their survey. In our
analysis, we determined a period of 5.644 days for this Cepheid which is close to the later period. 
We therefore considered only those Cepheids corresponds to our identified Cepheids where
two periods match within a day in two surveys. We note here that it is possible for POINT-AGAPE
to have two variables detected in such a close proximity with a completely different periods due
to their relatively deeper photometry and better sky conditions in La Palma.
In columns 8 and 9 of Table 1, we list the ID and period of the stars
found in POINT-AGAPE survey which closely matches with the Cepheids found in the present study.
Most of the periods agree within 0.06 days of each other except NMS-M31V25 which has a period
difference of 0.269 days.

\subsection{Frequency-Period distribution of the Cepheids}
The number of Cepheids observed in a galaxy is not uniformly distributed over all possible
periods.
The number of Cepheids occurring a certain range of
periods in a given complete sample depends on the initial mass function, chemical composition
of the galaxy, and the structure and evolutionary time scales of stars of different masses
during their transit through the instability strip. The frequency-period distribution
for classical Cepheids has been studied in detail by Becker, Iben \& Tuggle (1977),
Serrano (1983) and Alcock et al. (1999) for different galaxies and their studies show that
the frequency-period distribution is a function of chemical composition. Serrano (1983) also
pointed out that the mean period of Cepheids decreases with the galactocentric distance. Two
different explanations have been given to explain the bimodal pattern in frequency-period
distribution of Cepheids and deficiency of 8-10 days period by Becker, Iben \& Tuggle (1977)
and Boucher, Goupil \& Piciullo (1997). While former noted that it is two-component birth-rate
function responsible for the double peak, later suggested that it is in fact non-linear
fundamental pulsation cycle in 8-10 days range where corresponding Cepheids pulsate in the
first overtone having period $P_1 \approx 0.7 P_0$, resulting in an overall increase in overtone
Cepheids in the period range 5.6-7.0 days period which in turn shows a double peak in the
frequency-period distribution. Recently Antonello et al. (2002) made an extensive study of
the frequency-period distribution of 6 local group of galaxies and demonstrated that
while 3 metal-poor galaxies i.e. LMC, SMC and IC1613, do not show any conspicuous presence
of bimodal distribution in the frequency-period diagram, other 3 metal-rich galaxies i.e.
Milky Way, M31, and M33 have a visibly seen bimodal distribution.

\begin{figure}
\centering
\includegraphics[width=9.0cm,height=8.0cm]{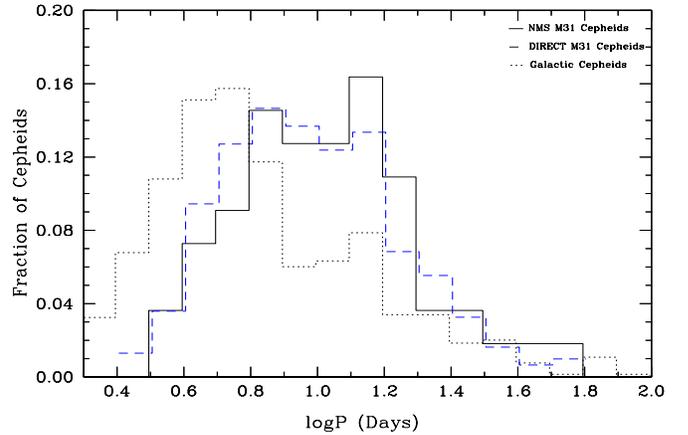}
\vspace{-2.3cm}
\caption{The normalized frequency-period distribution of Cepheids as a function of their pulsation
period in logarithmic scale. Cepheids identified in our target field detected under Nainital
Microlensing Survey survey and DIRECT survey are shown by solid and dashed lines while
Milky Way Cepheids taken from the GCVS catalogue are plotted by the dotted line. The
adopted bin size in all the 3 distributions is 0.1 in logP.}
\label{period_distri}
\end{figure}

To understand period distribution of Cepheids in the present study, we compared it with the
DIRECT survey (Macri 2004 and references theirin) where we found 332 Cepheids
in six of their targeted M31 field, 25 of them were repetitions among different fields.
We determined the fractional distribution of Cepheids against $logP$ for the two
catalogues--our catalogue of 55 Cepheids detected as a whole in the Nainital Microlensing
Survey and 307 Cepheids detected in the DIRECT survey. We estimated frequency of Cepheids
in each 0.1 bin of $logP$ in both the cases and normalized them with the total number of
Cepheids detected in the respective catalogues. In Fig.~\ref{period_distri}, we plot
histograms of these distributions. Though our conclusion
is based on a fairly small sample of Cepheids observed in a small region of M31 disk within
our survey and subject to statistical error but it clearly shows that period distribution
of Cepheids observed in M31 shows a bimodal pattern with peaks around $logP \sim 0.9$ and
1.1 days, a similar pattern as shown by the DIRECT Cepheids. However, Antonello et al. (2002)
found primary and secondary maxima at $logP \sim 0.7$ and 1.1 respectively 
using the sample of M31 Cepheids reported in the GCVS catalogue while Vilardell et al. (2007)
found these peaks close to 0.6 and 1.1 in their survey. On comparing the M31 Cepheids
frequency-period distribution with that of the Milky Way, we observed a bimodal period
distribution using 648 Galactic Cepheids reported in the GCVS catalogue with
peaks around $logP \sim 0.7$ and 1.1 days (see, Fig.~\ref{period_distri}), a pattern also
seen by Vilardell et al. (2007) who used the David Dunlop Observatory catalogue of Galactic
Cepheids (Fernie et al. 1995).

It is quite evident that the frequency-period distribution in M31 vary in shape and
in the location of the peak among different surveys. This is possibly due to
incompleteness of the Cepheids detected within these surveys. While GCVS and DIRECT surveys
seem incomplete towards shorter period, Vilardell et al. (2007) has attributed to the
observational biases for the long period Cepheids in their data. Apart from the limiting
magnitude of each surveys, these surveys are biased in their detection of Cepheids due
to non-uniform sampling of the data in different regions of the M31, non-detection
of low-amplitude Cepheids and blending of Cepheids by either foreground stars or nearby
bright stars within the host galaxy itself. To find a better understanding of the period
distributions of the Cepheids in M31, a systematic and deep photometric search of the
galaxy is needed in order to obtain a homogeneous data sample, at least in $BVI$ bands.

\subsection{Period-luminosity relation}
\begin{figure}
\centering
\includegraphics[width=9.0cm,height=8.0cm]{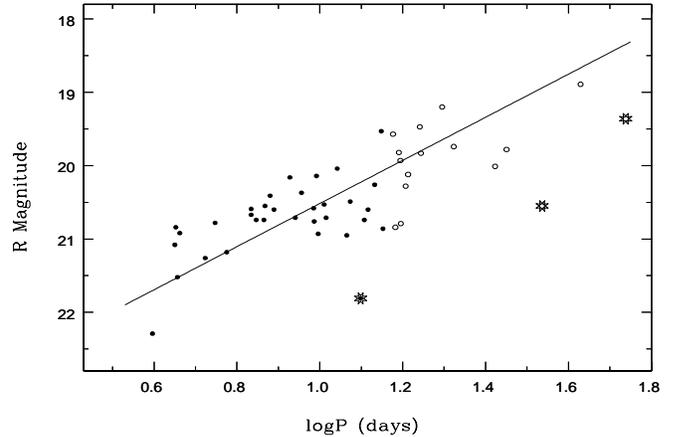}
\vspace{-2.3cm}
\caption{$R$ band Period-luminosity diagram for 49 Cepheids for which we have mean $R$ magnitude
available (see Table 1). Filled circles indicate short-period Cepheids identified in the present study
and open circles indicate the Cepheids with $P > 15$ days identified in the previous
study. The slope of the fitted straight lines are fixed at dm/d log$P$ = $-2.94$, given
by Madore \& Freedman (1991). 3 Cepehids marked with asterisk could be possible population
II Cepheids.}
\label{period_lumin}
\end{figure}

The Cepheid variables exhibit an excellent correlation between their mean intrinsic
brightness and pulsation period and widely used as a standard candles for estimating
extragalactic distances by comparing their absolute magnitudes inferred from
period-luminosity relation with their observed apparent magnitudes.
However, as we discussed earlier, short period and low-amplitude Cepheids are expected to be
affected by blending and hence the error in their mean magnitude is possibly dominated
by this as opposed to the  photometric error itself.  JOS03
reported detection of 26 Cepheids and 10 of these are detected in this study.
A period-luminosity diagram of the Cepheids detected in our survey is shown in
Fig.~\ref{period_lumin}. Here, we used all 55 Cepheids detected in our survey except
6 Cepheids for which we could not determined their $R$ magnitude.  We note that M31
Cepheids identified in our target field could have  more than 30\% inaccuracy in their
reported magnitudes due to combined effect of photometric error, errors in  flux
correlation and blending effect etc. Further, while Vilardell et al. (2007) suggested
not to use Cepheids below 0.8 mag amplitude in $V$ band in deriving the precise
distance of M31, we have chosen not to implement any such criteria in the present
study owing to large number of low amplitude Cepheids detected in our survey.

To estimate the distance of M31 from the $R$ band period-luminosity diagram, we kept the slope
and zero point fixed at -2.94, -4.52 respectively (Madore \& Freedman 1991) and used a total
extinction of 0.63 in our observed direction (JOS03). We have excluded 3 Cepheids from our
sample for the distance estimation which could be Population II objects as they
fall about 1.5 to 2 mag below the period-luminosity relation for the classical Cepheids
(indicated by an asterisk in Fig.~\ref{period_lumin}). We determined a distance modulus of
$(m-M)_0 = 24.41\pm0.21\pm0.30$ mag for M31. Here the first error indicates uncertainty
in the zero point while second error indicates typical photometric error at the faintest
magnitude level in our $R$ band data. Though our distance estimate is consistent with those
previously found for M31 (e.g. Freedman et al. 2001, Brown et al. 2004, Vilardell et al. 2007),
we emphasize that distance estimation based on the M31 Cepheids identified in our survey is a
crude estimation and much more precise photometry of these Cepheids at multiple wavelengths
is needed to ascertain an accurate distance to M31.

\section{Summary}
The main aim of the `Nainital Microlensing Survey' was to search for microlensing events
in the direction of M31. However, the vast amount of data also enabled us to identify
a substantial number of variable stars and optical transients in the $13^{'} \times 13^{'}$
region of the M31 disk. The data in the present study was analysed using the pixel technique
which is commonly used to look to detect variability in the crowded fields and/or poor seeing
conditions.
In this study, we have given a catalogue of 39 short-period Cepheids in the disk of M31
which were found within the magnitude range $\sim$ 19.5-22.3 in $R$ band. A
large number of photometric observations carried out over the survey's  4 years duration, has 
allowed us to determine their periods, which was found in the range between
$\sim$ 3 to 15 days. We note that although the phase coverage of the Cepheids is uneven in
our survey, these light curves cover many cycles of periodic variations hence derived periods
are reliable and generally in good agreement with previously published values. A correlation
between PSF-fitted photometric magnitudes, whenever possible, and corresponding pixel fluxes
were used to calculate mean magnitude and amplitude of variability of these Cepheids in $R$ and
$I$ bands. 
Fourier analysis is often used to distinguish classical Cepheids from those pulsating in their
first-overtone (Vilardell et al. 2007 and references theirin), however, the photometric quality
of our data is not good enough to study the pulsation modes of these Cepheids.

It is quite evident that while the long-period Cepheids are well represented in most of
the surveys, short-period Cepheids are under-sampled due to their low-amplitude, low
intrinsic brightness and poor data sampling. It is also difficult to detect short-period
Cepheids in the crowded field like M31 where the faint stars, particularly in poor seeing
conditions are visually undetectable from the background flux. However, it is demonstrated
in the present study that a much more complete sample of Cepheids can be obtained, even
among short-period and intrinsically faint Cepheids, using pixel technique. Our observation
of bimodal frequency-period distribution in a sample of 55 M31 Cepheids detected in the
Nainital Microlensing Survey is in agreement with such a trend seen by the other surveys,
however, a systematic search for the Cepheids is required to fully understand the
underlying reasons for the variations in shape and location of the peak in the frequency-period
distributions among different surveys. Due to observing limitations,
present sample does not contain stars with P shorter than 3.4 days and large uncertainties
in our magnitudes do not allow us to compare our results with the other galaxies on the
basis of their metallicities.

The growing number of Cepheids in distant galaxies are not only useful to determine their
precise distances but more so to trace star formation history of the galaxy itself.
Although our field of view is small in comparison of some other wide-field surveys
carried out in M31, our catalogue of short-period Cepheids detected through pixel method
bring a significant contribution towards lower branch of the period-luminosity diagram
and our results show that despite an average quality data, we can get comparative results
with the other surveys.

\begin{acknowledgements}
It is a great pleasure to thank Yannick Giraud-H\'{e}raud and Jean Kaplan for
their kind support in initiating the project and helping us to get familiar with
the pixel technique. We are thankful to Don Pollacco for his useful comments
on the initial draft of this paper and supporting staff at the Nainital observatory
for their help in carrying out sucessfull observations for such a long period.
This study is a part of the project 2404-3 supported by
the Indo-French center for the Promotion of Advanced Research, New Delhi.
\end{acknowledgements}

\label{lastpage}
\end{document}